\documentclass[traditabstract,longauth]{aa} 

\usepackage{graphicx}
\usepackage{txfonts}
\usepackage{subfigure}
\usepackage[authoryear]{natbib}
\usepackage{color}
\usepackage{url}
\usepackage{sidecap}
\usepackage{psfrag}
\begin{document}
   \title{The Vega Debris Disc: A view from Herschel\thanks{Herschel is an ESA space observatory with science instruments provided by European-led Principal Investigator consortia and with important participation from NASA.}}
   \subtitle{}
\author{
B.\,Sibthorpe\inst{1},
B.\,Vandenbussche\inst{2},
J.\,S.\,Greaves\inst{3},
E.\,Pantin\inst{4},
G.\,Olofsson\inst{5},
B.\,Acke\inst{2},
M.\,J.\,Barlow\inst{6},
J.\,A.\,D.\,L.\,Blommaert\inst{2},
J.\,Bouwman\inst{7},
A.\,Brandeker\inst{5},
M.\,Cohen\inst{8},
W.\,De\,Meester\inst{2},
W.\,R.\,F.\,Dent\inst{9},
J.\,Di\,Francesco\inst{10},
C.\,Dominik\inst{11,12},
M.\,Fridlund\inst{13},
W.\,K.\,Gear\inst{14},
A.\,M.\,Glauser\inst{15,1},
H.\,L.\,Gomez\inst{14},
P.\,C.\,Hargrave\inst{14},
P.\,M.\,Harvey\inst{16,17},
Th.\,Henning\inst{7},
A.\,M.\,Heras\inst{13},
M.\,R.\,Hogerheijde\inst{18},
W.\,S.\,Holland\inst{1},
R.\,J.\,Ivison\inst{1,19},
S.\,J.\,Leeks\inst{20},
T.\,L.\,Lim\inst{20},
R.~Liseau\inst{21},
B.\,C.\,Matthews\inst{10},
D.\,A.\,Naylor\inst{22},
G.\,L.\,Pilbratt\inst{13},
E.\,T.\,Polehampton\inst{20,22},
S.\,Regibo\inst{2},
P.\,Royer\inst{2},
A.\,Sicilia-Aguilar\inst{7},
B.\,M.\,Swinyard\inst{20},
C.\,Waelkens\inst{2},
H.\,J.\,Walker\inst{20},
R.\,Wesson\inst{6}
}
\institute{UK Astronomy Technology Centre, Royal Observatory Edinburgh, Blackford Hill, EH9 3HJ, UK\\
\email{bruce.sibthorpe@stfc.ac.uk}
\and
Instituut\ voor\ Sterrenkunde,\ Katholieke\ Universiteit\ Leuven,\ Celestijnenlaan\ 200\ D,\ B-3001\ Leuven,\ Belgium
\and
School\ of\ Physics\ and\ Astronomy,\ University\ of\ St\ Andrews,\ North\ Haugh,\ St\ Andrews,\ Fife\ KY16\ 9SS,\ UK
\and
Laboratoire\ AIM,\ CEA/DSM-CNRS-Universit\'e\ Paris\ Diderot,\ IRFU/Service\ d’Astrophysique,\ Bˆat.709,\ CEA-Saclay,\ 91191\ Gifsur-Yvette\ C´edex,\ France
\and
Deptartment\ of\ Astronomy,\ Stockholm\ University,\ AlbaNova\ University\ Center,\ Roslagstullsbacken\ 21,\ 10691\ Stockholm,\ Sweden
\and
Deptartment\ of\ Physics\ and\ Astronomy,\ University\ College\ London,\ Gower\ St,\ London\ WC1E\ 6BT,\ UK
\and
Max-Planck-Institut\ f\"ur Astronomie,\ K\"onigstuhl\ 17,\ D-69117\ Heidelberg,\ Germany
\and
Radio\ Astronomy\ Laboratory,\ University\ of\ California\ at\ Berkeley,\ CA\ 94720,\ USA
\and
ALMA\ JAO,\ Av.\ El\ Golf\ 40\ -\ Piso\ 18,\ Las\ Condes,\ Santiago,\ Chile
\and
National\ Research\ Council\ of\ Canada,\ Herzberg\ Institute\ of\ Astrophysics,\ 5071\ West\ Saanich\ Road,\ Victoria,\ BC,\ V9E\ 2E7,\ Canada
\and
Astronomical\ Institute\ Anton\ Pannekoek,\ University\ of\ Amsterdam,\ Kruislaan\ 403,\ 1098\ SJ\ Amsterdam,\ The\ Netherlands
\and
fdeling\ Sterrenkunde,\ Radboud\ Universiteit\ Nijmegen,\ Postbus\ 9010,\ 6500\ GL\ Nijmegen,\ The\ Netherlands
\and
ESA\ Research\ and\ Science\ Support\ Department\, Keplerlaan1,\ NL-2201 AZ,\ Noordwijk,\ The\ Netherlands
\and
School\ of\ Physics\ and\ Astronomy,\ Cardiff\ University,\ Queens\ Buildings\ The\ Parade,\ Cardiff\ CF24\ 3AA,\ UK
\and
Institute\ of\ Astronomy,\ ETH\ Zurich,\ 8093\ Zurich,\ Switzerland
\and
Department\ of\ Astronomy,\ University\ of\ Texas,\ 1\ University\ Station\ C1400,\ Austin,\ TX 78712,\ USA
\and
CASA,\,University\,of\,Colorado,\,389-UCB,\,Boulder,\,CO 80309,\,USA
\and
Leiden\ Observatory,\ Leiden\ University,\ PO\ Box\ 9513,\ 2300\ RA,\ Leiden,\ The Netherlands
\and
Institute\ for\ Astronomy,\ University\ of\ Edinburgh,\ Blackford\ Hill,\ Edinburgh\ EH9\ 3HJ,\ UK
\and
Space\ Science\ and\ Technology\ Department,\ Rutherford\ Appleton\ Laboratory,\ Oxfordshire,\ OX11\ 0QX,\ UK
\and
Department\ of\ Radio\ and\ Space\ Science,\ Chalmers\ University\ of\ Technology,\ Onsala\ Space\ Observatory,\ 439\ 92\ Onsala,\ Sweden
\and
Institute for Space Imaging Science,\ University\ of\ Lethbridge,\ Lethbridge,\ Alberta,\ T1J\ 1B1,\ Canada
}

\date{Received September DAY, YEAR; accepted MONTH DAT, YEAR}
\abstract{We present five band imaging of the Vega debris disc obtained using the \emph{Herschel Space Observatory}.  These data span a wavelength range of 70--500\,$\mu$m with full-width half-maximum angular resolutions of 5.6-36.9\arcsec.  The disc is well resolved in all bands, with the ring structure visible at 70 and 160\,$\mu$m.  Radial profiles of the disc surface brightness are produced, and a disc radius of 11\arcsec\ ($\sim 85$\,AU) is determined.  The disc is seen to have a smooth structure thoughout the entire wavelength range, suggesting that the disc is in a steady state, rather than being an ephemeral structure caused by the recent collision of two large planetesimals.}
\keywords{Stars: Vega -- Instrumentation: photometers -- Methods: observational}
\authorrunning{Sibthorpe et al. 2010}
\titlerunning{Herschel observations of the Vega debris disc}
\maketitle
\section{Introduction}
Debris discs, of which the \object{Vega} ($\alpha$ Lyrae) disc is the archetype, are characterised as discs of dusty material generated by the collision of planetesimals in belts surrounding main sequence stars.  The ages of the stars which exhibit these discs ($\sim$ 350\,Myr in the case of Vega; \citealt{Song2000}) precludes the possibility for this dust to be primordial, as the time scale to remove such dust is $\la$10\,Myr \citep{Backman1993,Wyatt2008}.

The debris disc around Vega was first detected by \cite{Aumann1984} as an infrared excess using the Infrared Astronomical Satellite (\emph{IRAS}; \citealt{IRAS1984}), and has been extensively studied in the infrared and submillimetre over the subsequent 25 years \citep[e.g.][]{Holland1998, Wilner2002, Su2005, Marsh2006}.  The appearance of the disc has been found to vary significantly across this wavelength regime, changing from a smooth axisymmetric structure in the infrared \citep[hereafter S05]{Su2005}, to a structure in the submillimetre, wherein the majority of the emission lies in two discrete clumps \citep{Holland1998}.

In order to understand the reason for the variation in structure with wavelength it is important to first understand the origin of the clumps seen in the submillimetre.  The recent collision of two massive planetesimals is one option, however, given the age of Vega, the statistical likelihood of this occurring with two bodies of sufficient mass to explain the submillimetre observations is low \citep{Wyatt2002}.  A more favourable alternative, first proposed by \cite{Wilner2002} and modelled by \citet[hereafter W06]{Wyatt2006} and \cite{Reche2008}, is that the clumps are dust grains trapped in resonance with a planet near to the disc.  In this scenario the large dust grains (larger than a few mm) are trapped in these resonances, while smaller intermediate sized grains (a few $\mu$m--mm), having been perturbed by radiation pressure, have a more uniform distribution in the disc.

Recent analysis and modelling of \emph{Spitzer} \citep{Werner2004} mid-infrared data have reached contradictory conclusions. S05 find the disc to be ephemeral; in this scenario  the disc is the result of a recent massive collision of planetesimals, and the subsequent collisional cascade.  This results in a high mass of very small grains (less than a few $\mu$m) which are blown out of the system by radiation pressure immediately upon creation, resulting in the large disc extent observed.  Conversely, \cite{Muller2010} succeed in reproducing the surface brightness radial profile using intermediate size grains in elliptical orbits around the parent planetesimal ring, and therefore conclude that it is consistent with a steady-state model.  In the steady-state model, dust that is destroyed, either by being drawn in to the star due to Poynting-Robertson drag or blown out of the disc by radiation pressure, is continuously replenished by a steady collisional cascade within the planetesimal belt. 

If the small blown-out grains are the origin of the emission observed in the mid-infrared then W06 predicts that spiral features, emanating from the submillimetre clumps, should be visible with high-resolution imaging; a smooth structure would support the steady-state model.

In this paper we present five-band far-infrared imaging of the Vega debris disc obtained with the \emph{Herschel} \citep{Pilbratt2010} Photodetector Array Camera and Spectrometer \citep[PACS;][]{Poglitsch2010} and Spectral and Photometric Imaging Receiver \citep[SPIRE;][]{Griffin2010}.  We discuss the initial analysis and disc parameterisation, and relate these results to the ephemeral and steady-state disc models. In Section 2 we present the \emph{Herschel} data, outline the processing performed, and analyse the disc structure and properties.  These data are then compared with results from the recent \emph{Spitzer} observations (S05) and disc modelling of W06 and \cite{Muller2010}, with our conclusions summarised in Section 3.
\section{Observations and data processing}
We obtained images of Vega and its associated debris disc at 70 and 160\,$\mu$m with PACS, and 250, 350, and 500\,$\mu$m with SPIRE.  The data cover an angular scale of $\sim$25 and 64 sq. arcmin. for PACS and SPIRE respectively, with beam full-width half-maxima (FWHM) of $\sim$5.6, 11.3, 18.1, 25.2, 36.9\arcsec\ for the short to long wavelength bands.  The data were obtained in scan-map mode for both instruments using the nominal observing parameters in both cases, and scanning rates of 10 and 30\arcsec\ per second for PACS and SPIRE respectively.  The total on-sky observing time was 5506\,s and 6120\,s and comprised 70 and 16 map repetitions for PACS and SPIRE respectively.
\begin{figure*}
\begin{center}$
\begin{array}{ccccc}
\includegraphics[trim= -5mm 10mm 2mm  0mm, clip=true, scale=0.38]{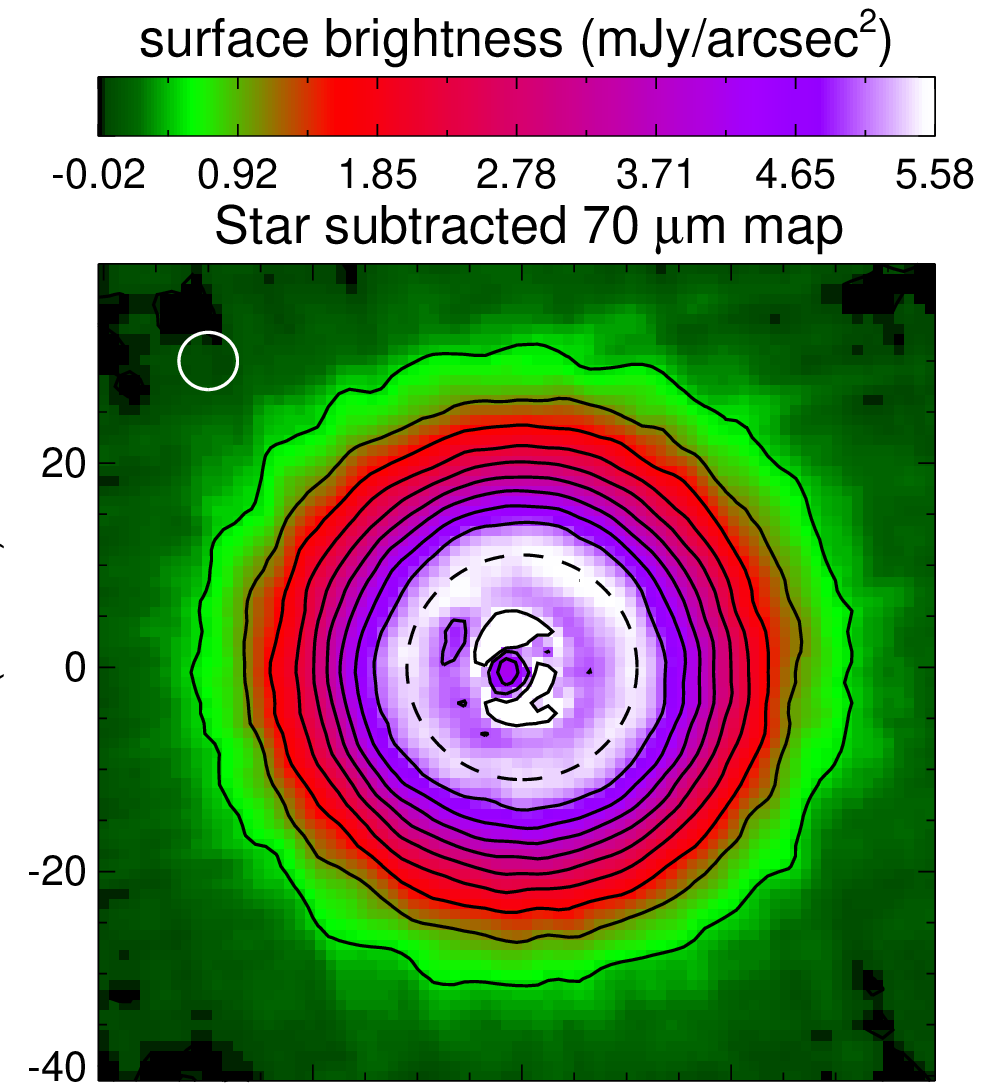}  &
\includegraphics[trim= 10mm 10mm 2mm  0mm, clip=true, scale=0.38]{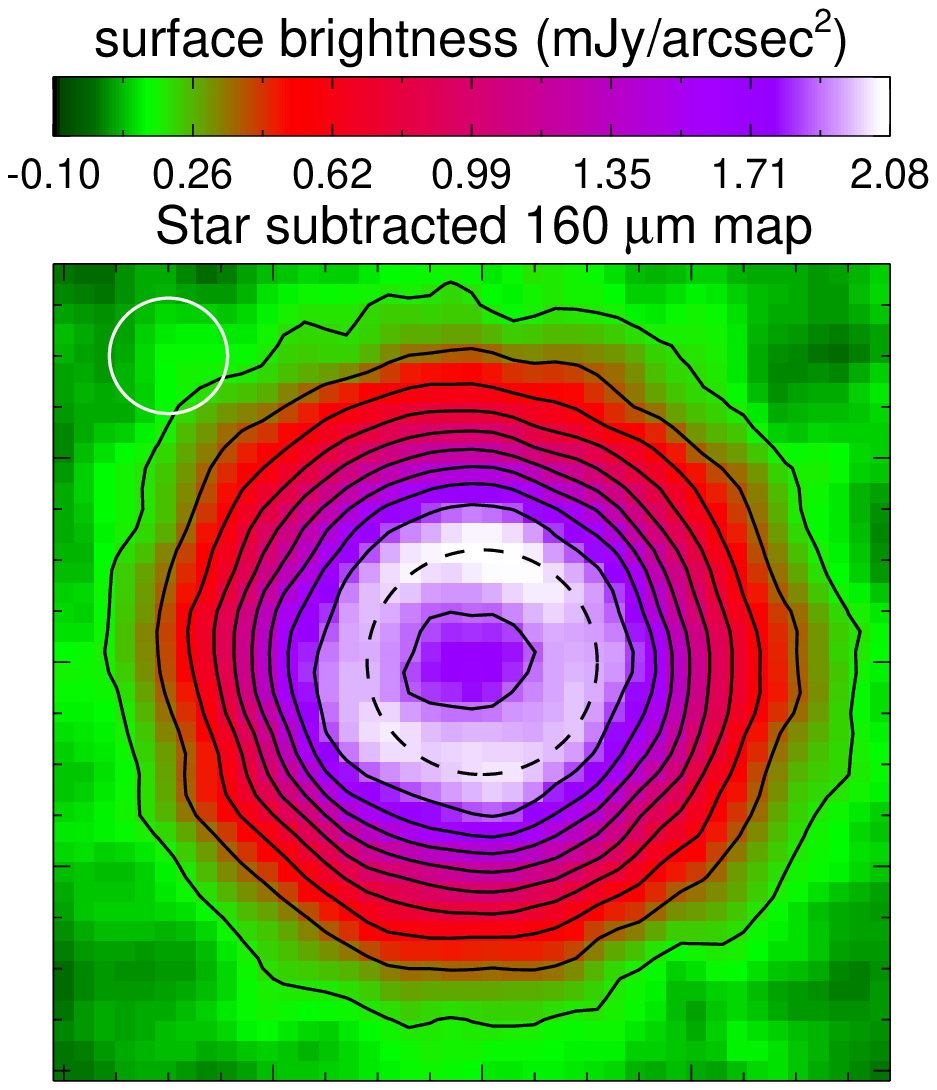}  &
\includegraphics[trim= 10mm 10mm 2mm  0mm, clip=true, scale=0.38]{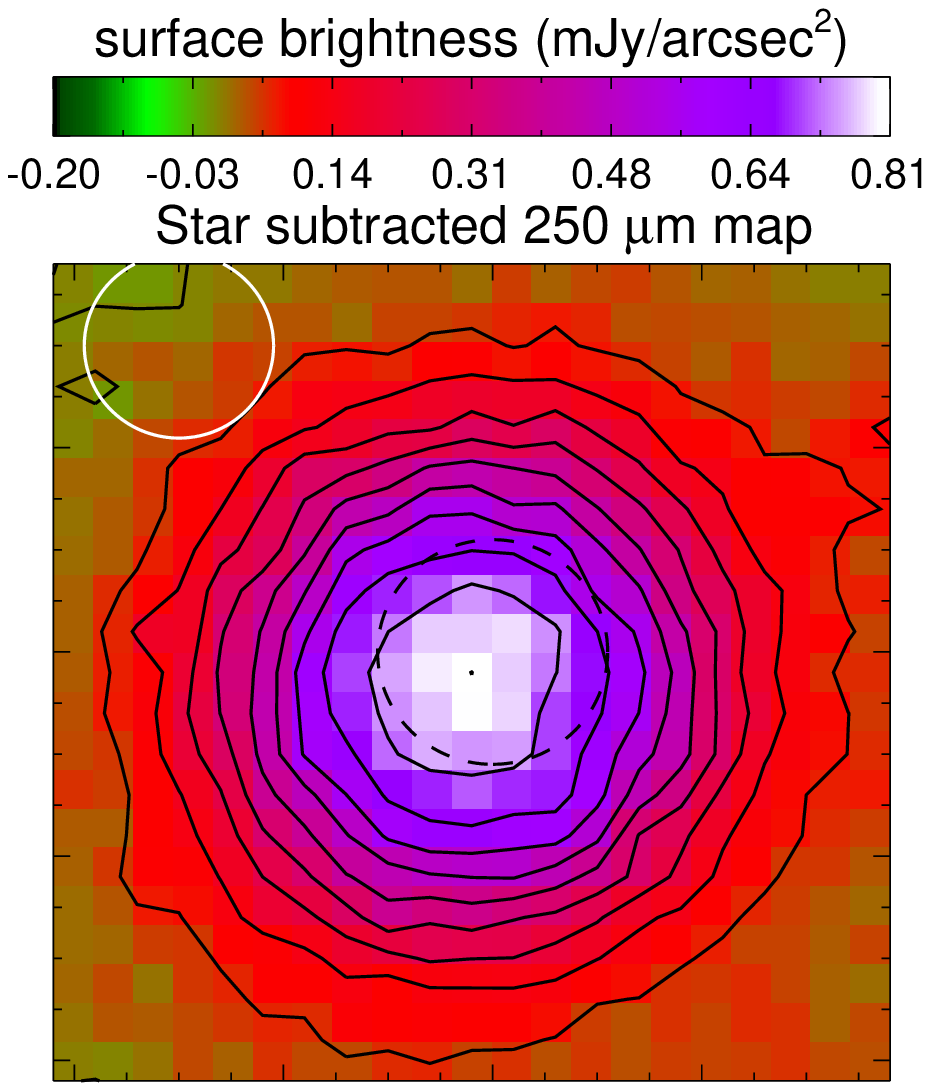}  &
\includegraphics[trim= 10mm 10mm 2mm  0mm, clip=true, scale=0.38]{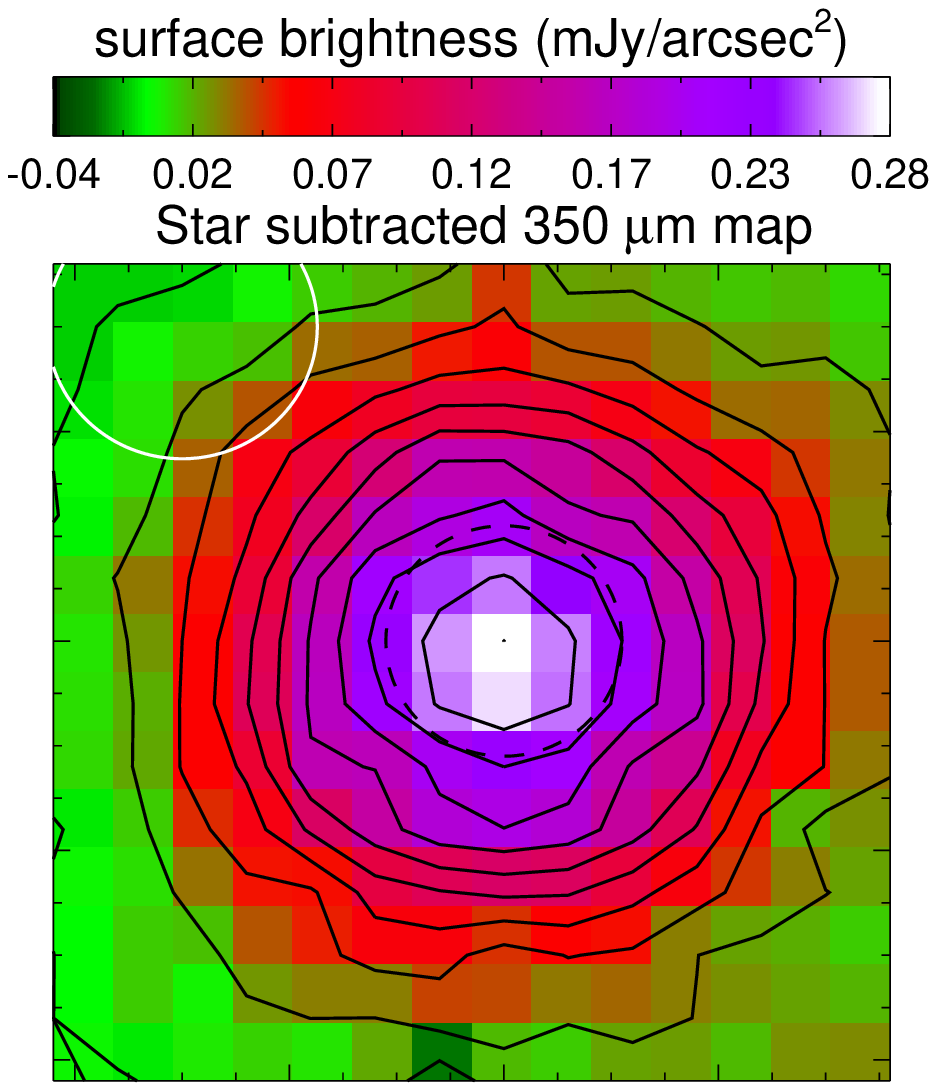}  &
\includegraphics[trim= 10mm 10mm 2mm  0mm, clip=true, scale=0.38]{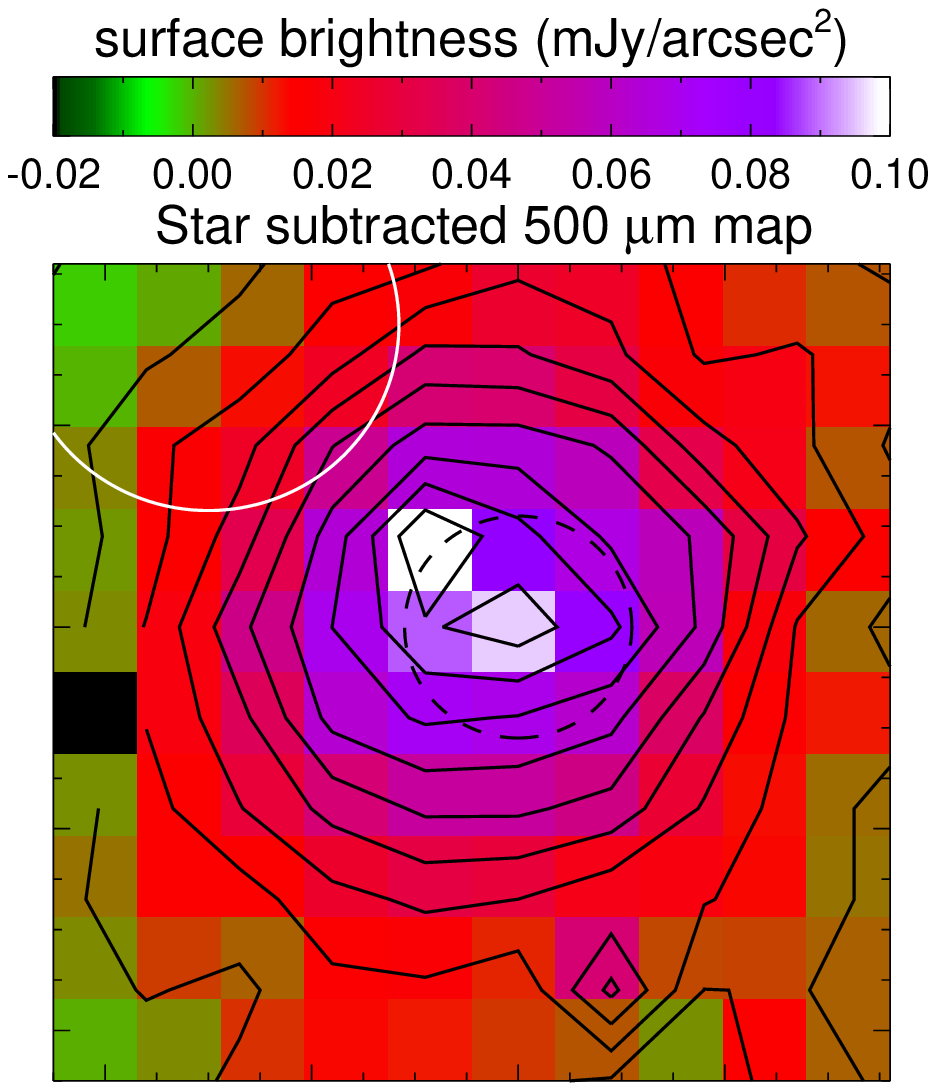}  \\

\includegraphics[trim= -5mm 0mm 2mm  19mm, clip=true, scale=0.38]{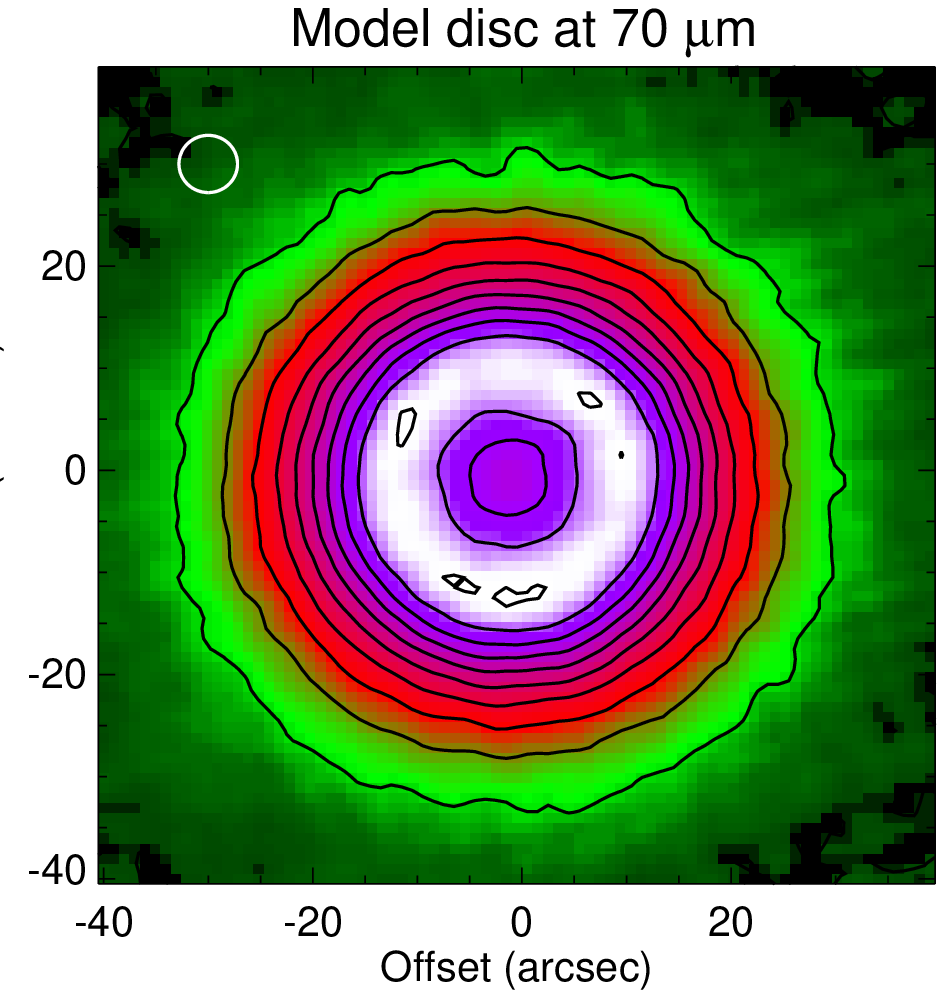}  &
\includegraphics[trim= 10mm 0mm 2mm  19mm, clip=true, scale=0.38]{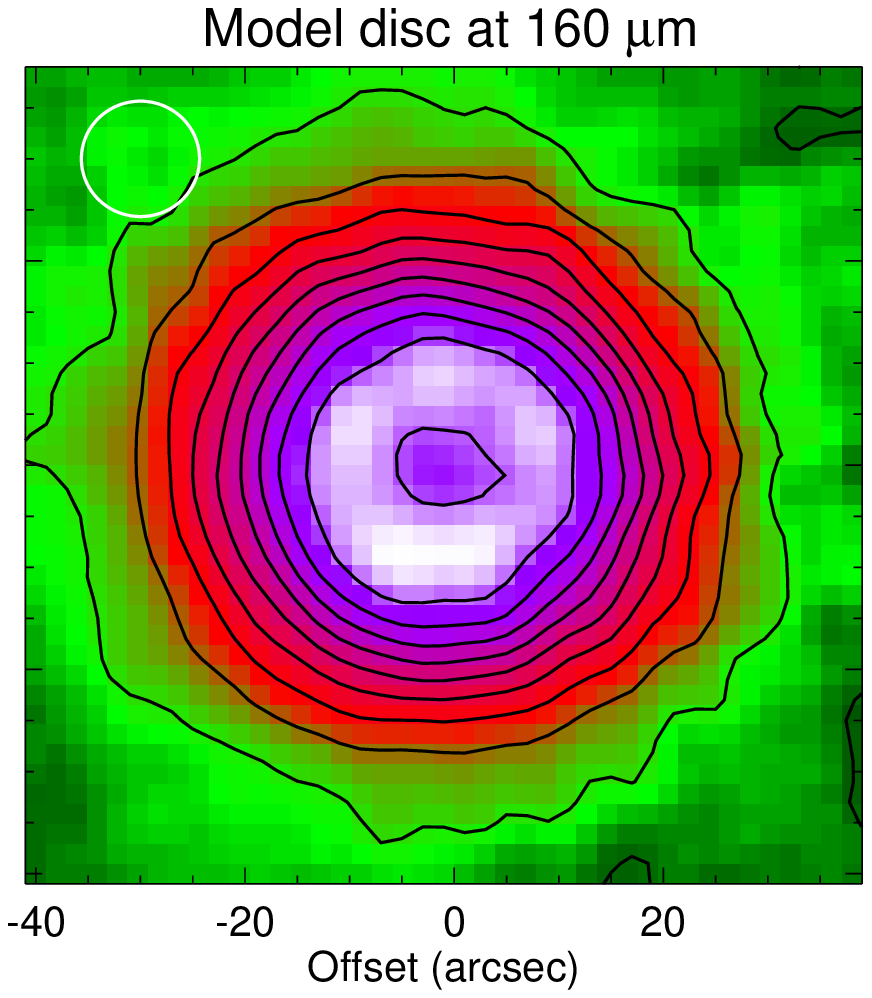}  &
\includegraphics[trim= 10mm 0mm 2mm  19mm, clip=true, scale=0.38]{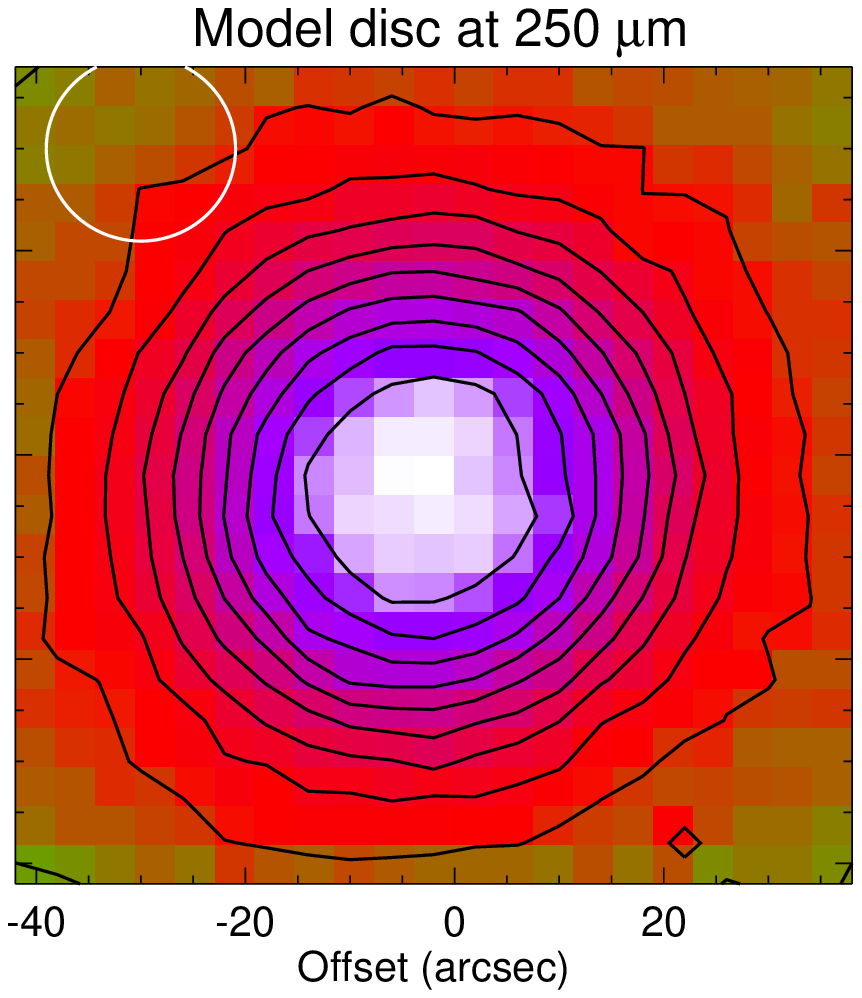}  &
\includegraphics[trim= 10mm 0mm 2mm  19mm, clip=true, scale=0.38]{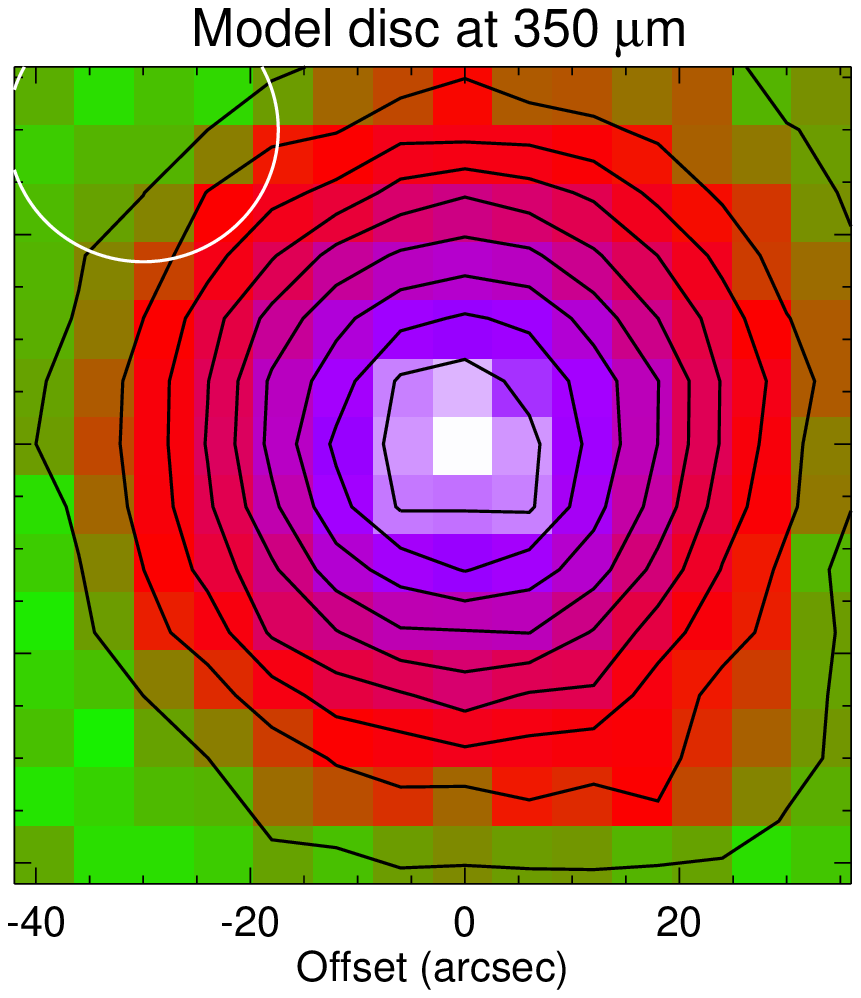}  &
\includegraphics[trim= 10mm 0mm 2mm  19mm, clip=true, scale=0.38]{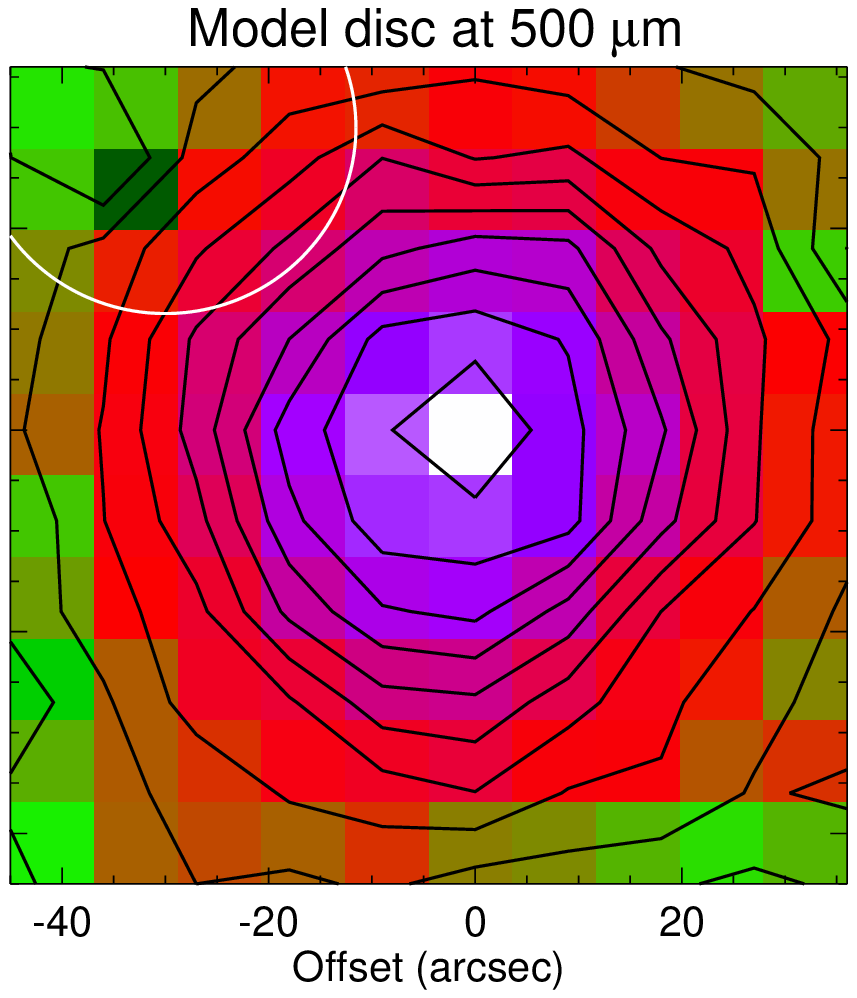}  \\
\end{array}$
\end{center}
\vspace{-5mm}
\caption{Data for the Vega debris disc from 70\,$\mu$m to 500\,$\mu$m from left to right respectively (top row - star-subtracted images; bottom row -  modelled images).  All images are scaled linearly, and both images within a given band are equally scaled.  The white circle represents beam FWHM in each band, and the contour lines are in steps of 5\% of the peak flux.  The black dashed circle represents the location of the disc at a radius of 11\arcsec.}
\label{maps}
\end{figure*}

The PACS data were high-pass filtered to remove low-frequency noise using a cut-off scale of 3.7\arcmin.  The data were then binned to a map using the default \emph{photProject} method in the \emph{Herschel} interactive processing environment \citep[HIPE;][]{Ott2010}.  The pixelisation of these maps was set to 1 and 2\arcsec\ per pixel, equivalent to $\sim$1/5 of a beam.  The maps have 1-$\sigma$ noise levels of 0.045 and 0.083\,mJy\,arcsec$^{-2}$, which includes 10 and 20\% flux calibration uncertainties, in the 70 and 160\,$\mu$m bands.  The background was removed from the maps by subtracting the median in the local vicinity of the source.

The SPIRE data were also reduced using HIPE and maps were obtained via the default \emph{na\"iveMapper} task.  The 16 repeat observations allowed the data to be binned to 4, 6, and 9\arcsec\ sized pixels without losing complete sampling across the source.  The 1-$\sigma$ noise level obtained was 0.014, $1.8\times10^{-4}$, and $2.8\times10^{-4}$\,mJy\,arcsec$^{-2}$ in the 250, 350, and 500\,$\mu$m bands respectively; the calibration error is $\sim$15\% \citep{Swinyard2010}.  The SPIRE beams exhibit a 1.07, 1.12, and 1.09 ellipticity, as described in the SPIRE beam model release note.

In order to more easily assess the structure in the disc, the photospheric contribution from the star was subtracted from the image (top row Fig.~\ref{maps}).  This was achieved by scaling a high signal-to-noise ratio observation of Vesta, which can be regarded as a point-source image, to an appropriate flux level to obtain a model for the stellar contribution in each band.  The Vesta image was rotated to match the position angle of the telescope used in the observation of Vega before subtraction of the model star.  The photospheric flux in each band was estimated using data given by \cite{Rieke2008}, and colour corrections of 1.02, 1.07, 0.96, 0.99 and 1.04 were applied for the 70 to 500\,$\mu$m bands respectively \citep{Poglitsch2010,SPIREOM}.  The resulting flux density estimates were 793, 162, 63, 32 and 16\,mJy for the 70 to 500\,$\mu$m bands respectively.  The stellar models were then subtracted from the reduced data.

The beam model used in star subtraction for both the PACS and SPIRE data was obtained using the same observing mode parameters as the original data.  The PACS beam exhibits a characteristic tri-lobe structure, while the SPIRE beam contains clear side-lobe structure and a small ellipticity.  Uncertainties in these beam models result in star subtraction artifacts in the central region of the image.  This is especially pronounced in the 70\,$\mu$m image where the stellar contribution is highest.  Consequently the star-subtracted maps can only be used to assess the disc structure at radii larger than 5 arcsec ($\sim$40 AU).
\subsection{Image analysis}
The five star-subtracted maps are presented in the top row of Fig.~\ref{maps}, and range from 70--500\,$\mu$m from left to right.  The disc is resolved in all bands, and shows a smooth and axisymmetric structure; centroids of the stellar and disc components show less than 1 pixel difference in the position of the star with respect to the disc in the 70 and 160\,$\mu$m bands, where the star and disc locations can be readily identified.  The 30\% contour of the five star-subtracted discs were fit by an ellipse and were found to be extremely circular, with ellipticities of $\sim$1.01$\pm$0.002, 1.02$\pm$0.003, 1.04$\pm$0.04, 1.03$\pm$0.06 and 1.11$\pm$0.09 for 70--500\,$\mu$m.  The 500\,$\mu$m image exhibits a significantly higher ellipticity than the other bands, however, the relatively low resolution in this band coupled with the 1.09 intrinsic beam ellipticity make the statistical significance of this measurement too low to draw a robust conclusion.  The flux densities, obtained via aperture photometry, for the star plus disc system, are 10.12$\pm$1.18, 4.61$\pm$0.9, 1.68$\pm$0.26, 0.61$\pm$0.10 and 0.21$\pm$0.04\,Jy from 70 to 500\,$\mu$m respectively, with the calibration error dominating the uncertainty.  These measurements agree well with integrated measurements made at similar wavelengths by other facilities \citep{Marsh2006,Su2006}.  A single aperture of radius 30\arcsec\ was used for all bands, with the same aperture randomly placed around the source to quantify the noise.

With the star-subtracted from the image the ring structure of the disc can be clearly identified at 70 and 160\,$\mu$m.  The disc is at a radius of $\sim$11\arcsec, which at the distance of Vega (7.76\,pc) corresponds to $\sim$85\,AU.  This equates to $\sim$4 beam half width half maxima from the central star, making this detection robust against artifacts from the star subtraction.  This is in agreement with previous infrared (S05), and submillimetre estimates \citep{Holland1998,Marsh2006}.  The inner cavity is not visible in the SPIRE data due to the decreased resolution relative to the shorter wavelength bands; the large scale disc size remains comparable.
\subsection{Radial profiles and surface brightness modelling}
The face-on nature of the Vega disc allows us to obtain data on the general disc structure and extent by azimuthally averaging the radial intensity profiles.  S05 performed such an analysis for the 24, 70, and 160\,$\mu$m data from \emph{Spitzer} and find that the disc profile can be fitted by $r^{-3}$ and $r^{-4}$ power laws for the inner and outer disc respectively.  Radial profiles for Vega, derived from the higher resolution \emph{Herschel} data, are presented in Fig.~2 for the raw and star-subtracted maps.  The stellar model used for star subtraction is plotted for reference.  Radial step sizes equal to the map pixel scale were used out to a radius of 90\arcsec.

The disc extends to a radius of $\sim$1\arcmin\ in all bands before the signal-to-noise ratio drops and the data become subject to uncertainties in the baseline removal.  The disc radius is in agreement with that found by S05 in both 70 and 160\,$\mu$m bands, with differences 6.4 and 2.8\% respectively.  To compare these data the PACS data were convolved to the \emph{Spitzer} resolution and the disc radius was measured at a surface brightness of 0.5\,mJy\,arcsec$^{2}$.  The PACS radii measured at the \emph{Spitzer} resolution were 34.7$\pm^{+0.9}_{-0.8}$\,\arcsec\ and 31.6$^{+3.5}_{-3.2}$\,\arcsec\ for the 70 and 160\,$\mu$m bands respectively, with the errors based on the flux calibration uncertainties.

The drop-off in radial profile at high radius appears linear in the log-linear plots in Fig.~2.  The functional form of this slope, characterised between radii of 20--50\arcsec, is ${\log_{10}}(S_{\nu})=-0.63r+1.61$ at 70\,$\mu$m, where $S_{\nu}$ is the surface-brightness at radius $r$.  This is in contrast to the power-law slope identified by S05 for the PACS bands.  However, S05 fit the disc out to larger radii, and within the region we fit the same functional form could be similarly applicable.  An accurate comparison with S05 at radii larger than $\sim$50\arcsec\ if difficult as these data are highly affected by uncertainties in the background subtraction.

The clearly defined ring identified in Fig.~\ref{maps} is evident again in the radial profiles at 70 and 160\,$\mu$m, and is defined by the peak and turn-over of the disc profile at a radius of 11\arcsec.  The structure observed, however, makes it difficult to obtain a discrete measurement of the inner and outer edges of the disc.  As an alternative characterisation we measured a half-width half-maximum (HWHM) size of the disc, outward in radius from the peak disc brightness at 11\arcsec\ (Fig. 2), and using this peak as the reference maximum.  We obtained HWHM sizes for the 70 and 160\,$\mu$m bands, in which the disc radius is identifiable, of 9 and 11.3\arcsec\ respectively.
\begin{figure}
\begin{center}
\includegraphics[trim= 6mm 13.5mm 0mm 00mm, clip=true, scale=0.5]{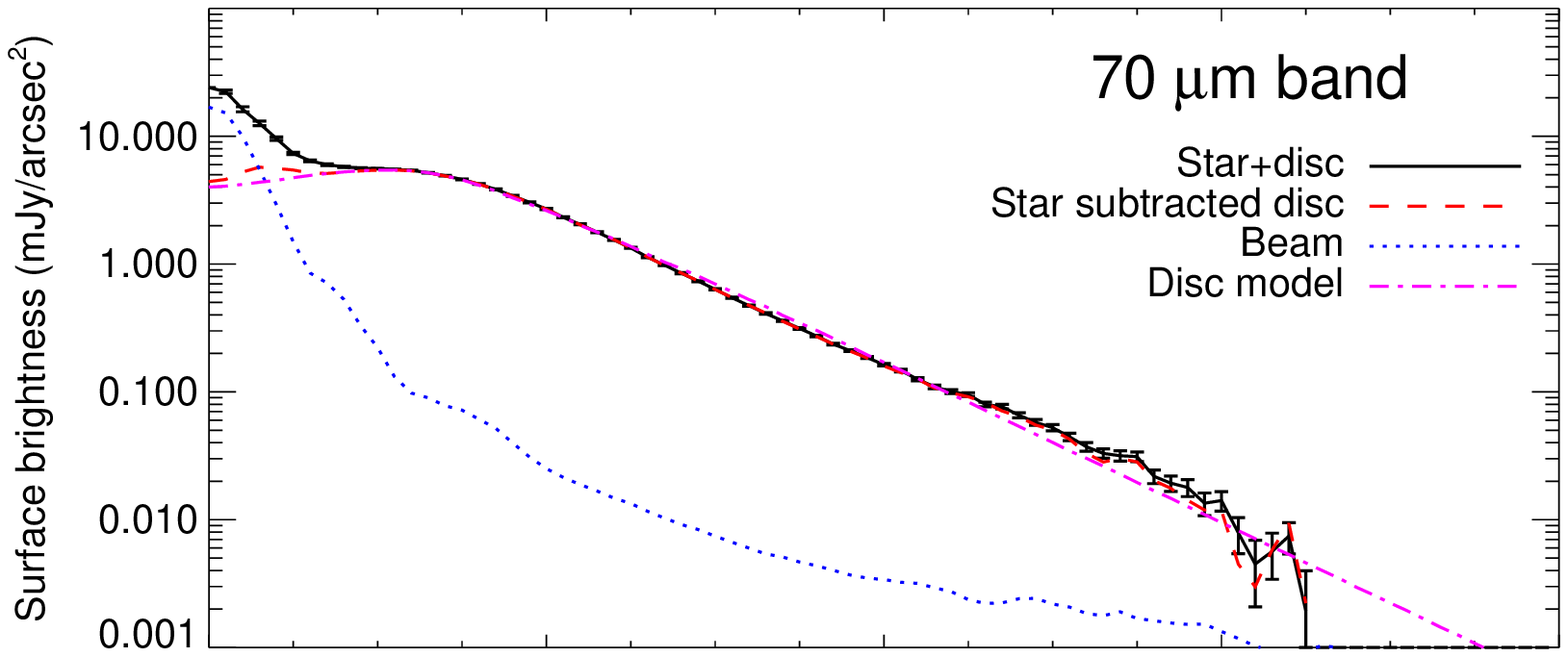}
\includegraphics[trim= 6mm 13.5mm 0mm 11mm, clip=true, scale=0.5]{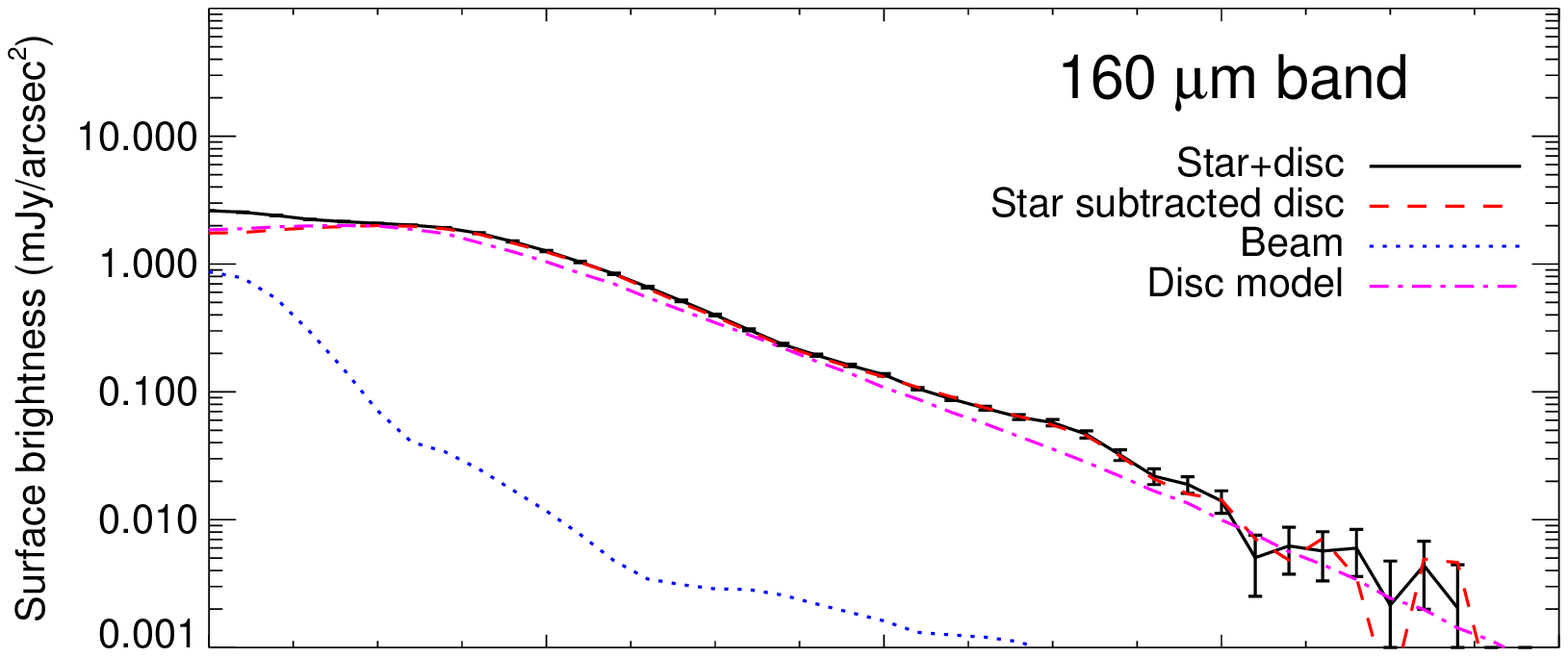}
\includegraphics[trim= 6mm 13.5mm 0mm 11mm, clip=true, scale=0.5]{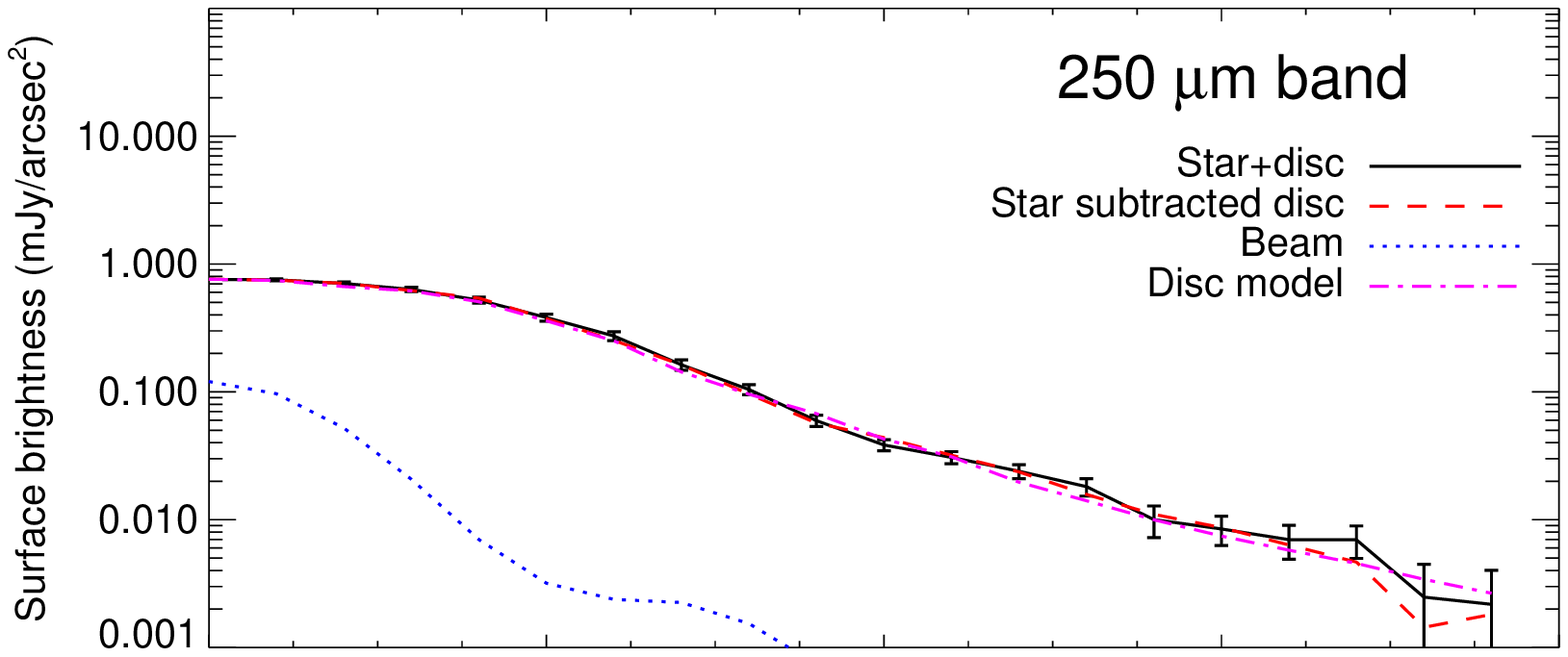}
\includegraphics[trim= 6mm 13.5mm 0mm 11mm, clip=true, scale=0.5]{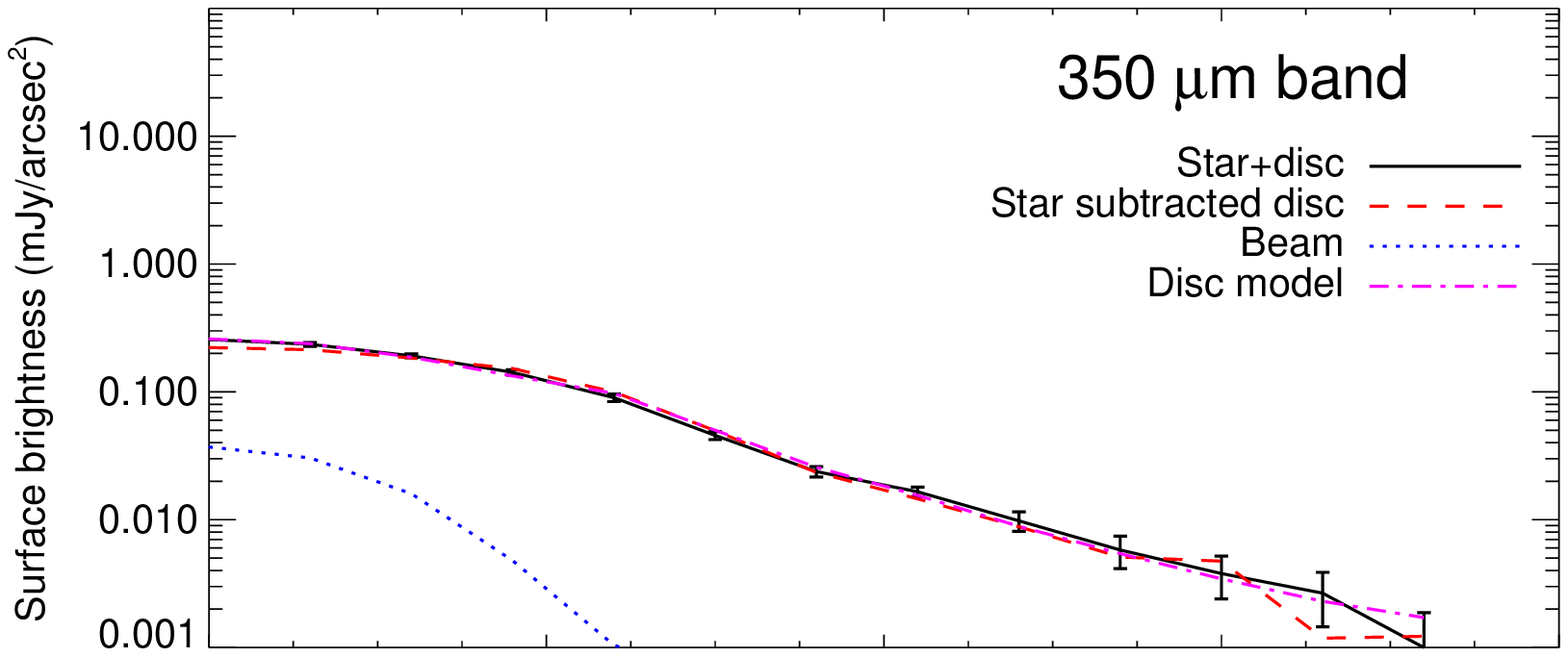}
\includegraphics[trim= 6mm 00.0mm 0mm 11mm, clip=true, scale=0.5]{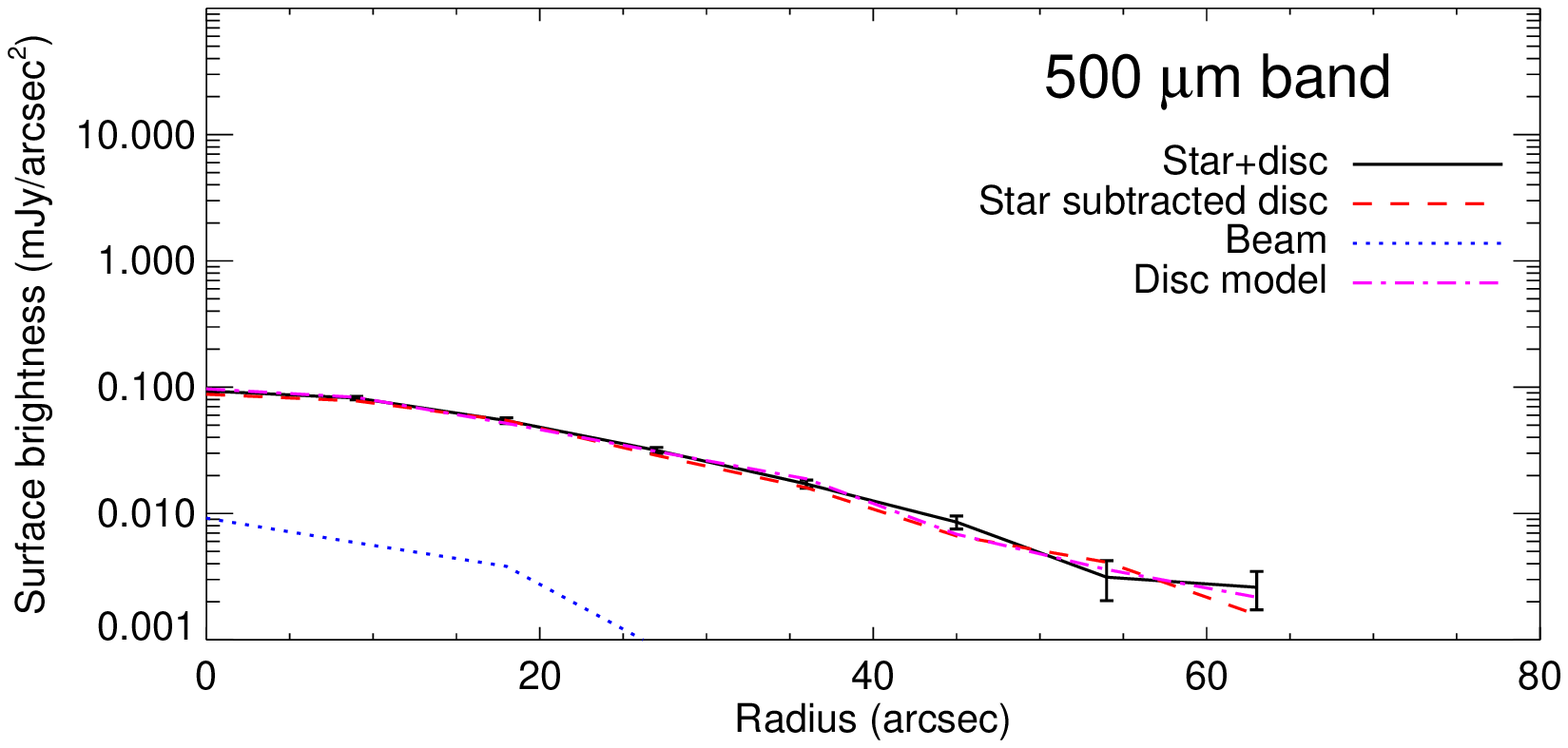}
\vspace{-2mm}
\caption{Radial Surface brightness profiles in the five \emph{Herschel} bands for the original, the star-subtracted, and modelled data, as well as the scaled stellar photospheric model. Error bars of 1-$\sigma$ are given for the star plus disc measurements.}
\end{center}
\vspace{-8mm}
\label{radProfile}
\end{figure}

To further assess the potential disc structure, we developed a simple model based only on the observed surface brightness properties at 70\,$\mu$m.  The model comprises two different surface brightness distributions, separated at a reference radius, $r_0$.  The inner profile, $r < r_{0}$ was chosen to be a Gaussian function peaking at $r_{0}$, and the outer profile, $r_{0} < r$, was defined using the functional form found for the 70\,$\mu$m outer disc described above.  The distinction between the inner and outer discs, as parameterised in this model by $r_{0}$, is a purely observational definition, based on the transition from one brightness distribution model to the other.  A more physically meaningful definition for the separation between the inner and outer disc is the peak of the disc profile found at a radius of 11\arcsec\ as described above.  The model was then convolved with the appropriate beam model and compared to the measured surface brightness profile.

A best-fit model was found with parameters $r_0=14$\arcsec\ and inner Gaussian FWHM = 20\arcsec.  Models for the bands longward of 70\,$\mu$m were created using the same intrinsic surface brightness profile and convolved with the appropriate beam.  The peak value of the resultant models were scaled by 0.33, 0.10, 0.04 and 0.01 for the 160 to 500\,$\mu$m bands respectively to match the data, and placed in an empty region of the original map to replicate the realistic instrumental noise (bottom row Fig.~\ref{maps}).  The radial profile of the output modelled image was measured and plotted with the real data in Fig.~2.  There is good agreement between the modelled and real radial profiles, with 1-$\sigma$ residuals in the 20-50\arcsec\ fitted region below 0.05\,mJy per sq. arcsec across all five bands, implying that the underlying structure of the disc across the wavelength range is similar.

As the model data are known to be perfectly smooth, a direct comparison can be made with the real data to assess the significance of any potential small scale structure seen within the disc.  For example, at 160\,$\mu$m very low-level structure can be seen, with flux enhancements in the northern and southern parts of the ring.  However, the counterpart model disc image in Fig.~\ref{maps} shows similar features, indicating that these are at the level of the noise, and should not be attributed to true disc structure.

\begin{figure}
\centering
\includegraphics[width=0.3\textwidth]{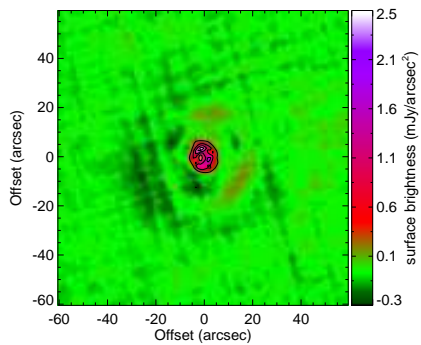}
\caption{Difference image for the star-subtracted data and the model image at 70\,$\mu$m.  In this instance the model image is noiseless.}
\vspace{-3.5mm}
\label{modelDiff}
\end{figure}

Subtraction of this simple uniform model, without added noise, from the original data provides another method to easily identify disc structure which is otherwise difficult to detect in the presence of the larger disc; a difference map at 70\,$\mu$m, the highest resolution band, is shown in Fig.~\ref{modelDiff}.  The brightest structure in this image lies within the inner disc, where the beam subtraction artifacts are strongest.  This can also be seen in Fig.~2 as a difference between the radial profiles of the model and the star-subtracted image at $\sim$6\arcsec.  The dark features seen in the outer disc region also correspond in structure and position to the triple-lobed beam pattern, and therefore are disregarded as potential disc features.  With the exception of the beam subtraction structures, there is no sign of any other clumpy structure associated with the disc, down to the noise limit of these data.

\section{Discussion and conclusions}
The structure observed in the \emph{Herschel} data shows no sign of clumps. There are also no visible spiral arm features, predicted by W06, if the disc emission at mid-to far-infrared wavelengths is dominated by small blown-out dust grains (W06 Fig.~3; right hand panels, $\beta=$1--10).  The smooth structure observed is most consistent with the steady-state model, wherein the emission is dominated by intermediate size dust grains in elliptical orbits about the parent planetesimal belt.  This model was found to simultaneously give good agreement to the data in all bands, which is unexpected, as the more distant grains should have a lower temperature, and suggests that the mean grain size decreases with distance from the star.  This is in-keeping with observational data which shows a larger disc at shorter wavelengths.  Full modelling of the radial grain size distribution will be presented in Sibthorpe (2010, in prep).
\begin{enumerate}
   \item We presented resolved images of the Vega debris disc system in five bands ranging from 70--500\,$\mu$m obtained using the \emph{Herschel} PACS and SPIRE instruments.
   \item The peak surface brightness of the dust disc was identified at 70 and 160\,$\mu$m at a radius of 11\arcsec\ (85 AU).
   \item The surface brightness profile was found to be well fit in the outer disc by a $\log_{10}(S_{\nu}) \propto -0.63r$ distribution, with a different scale factor at each band.  The inner profile ($r \le r_{0}$) was likewise modelled, with a Gaussian profile of FWHM = 20\arcsec\ found to provide a good fit.  The change in surface brightness distribution, occuring at a radius of $\sim$14\arcsec\ ($\sim$109 AU), is used to observationally define the distinction between the inner and outer disc. This model was found to simultaneously give good agreement to the data in all bands.
   \item The structure of the disc was found to be smooth, with no clumpy structure to the sensitivity limit of these data.
   \item While these data cannot preclude the option that the Vega disc is the result of a large planetesimal collision, making it ephemeral in nature, these data support the hypothesis that the Vega disc is steady-state in nature.
\end{enumerate}
\begin{acknowledgements}
The authors would like to acknowledge the PACS and SPIRE instrument teams. PACS has been developed by a consortium of institutes led by MPE (Germany) and including UVIE (Austria); KU Leuven, CSL, IMEC (Belgium); CEA, LAM (France); MPIA (Germany); INAFIFSI/OAA/OAP/OAT, LENS, SISSA (Italy); IAC (Spain). This development has been supported by the funding agencies BMVIT (Austria), ESA-PRODEX (Belgium), CEA/CNES (France), DLR (Germany), ASI/INAF (Italy), and CICYT/MCYT (Spain). SPIRE has been developed by a consortium of institutes led by Cardiff Univ. (UK) and including Univ. Lethbridge (Canada); NAOC (China); CEA, LAM (France); IFSI, Univ. Padua (Italy); IAC (Spain); Stockholm Observatory (Sweden); Imperial College London, RAL, UCL-MSSL, UKATC, Univ. Sussex (UK); Caltech, JPL, NHSC, Univ. Colorado (USA). This development has been supported by national funding agencies: CSA (Canada); NAOC (China); CEA, CNES, CNRS (France); ASI (Italy); MCINN (Spain); SNSB (Sweden); STFC (UK); and NASA (USA).  BA is supported by postdoctoral fellowship funding from the Fund for Scientific Research, Flanders.
\end{acknowledgements}

\bibliographystyle{aa}
\bibliography{references}

\end{document}